\documentclass[12pt,a4paper]{article}
\pdfoutput=1

  \usepackage{a4wide}
  \usepackage{latexsym}
  \usepackage{epsf}
  \usepackage{amssymb}
  \usepackage{graphicx}
  \usepackage{amsmath, cite}
 \usepackage{bbm}
  \usepackage{amsmath,amssymb,amsthm}
  \usepackage{verbatim}
  \usepackage{hyperref}
\renewcommand{\d}{\textrm{d}}
\newcommand{\e}{\textrm{e}}

\newcommand{\w}{\wedge}

\newcommand{\SO}{\mathop{\rm SO}}
\newcommand{\SU}{\mathop{\rm SU}}
\newcommand{\be}{\begin{equation}}
\newcommand{\ee}{\end{equation}}
\newcommand{\ba}{\begin{eqnarray}}
\newcommand{\ea}{\end{eqnarray}}

\newcommand{\lp}{\left(}
\newcommand{\rp}{\right)}

\newcommand{\sgn}{\textrm{sgn}\,}

\newcommand\varpm{\mathbin{\vcenter{\hbox{%
  \oalign{\hfil$\scriptstyle+$\hfil\cr
          \noalign{\kern-.3ex}
          $\scriptscriptstyle({-})$\cr}%
}}}}

\newcommand\varmp{\mathbin{\vcenter{\hbox{%
   \oalign{\hfil$\scriptstyle-$\hfil\cr
           \noalign{\kern-.3ex}
          $\scriptscriptstyle({+})$\cr}%
}}}}

\newcommand\varbs{\mathbin{\vcenter{\hbox{%
   \oalign{\hfil$\scriptstyle>$\hfil\cr
           \noalign{\kern-.3ex}
          $\scriptscriptstyle({<})$\cr}%
}}}}

\newcommand\varsb{\mathbin{\vcenter{\hbox{%
   \oalign{\hfil$\scriptstyle<$\hfil\cr
           \noalign{\kern-.3ex}
          $\scriptscriptstyle({>})$\cr}%
}}}}

\begin{document}
\numberwithin{equation}{section}
\begin{flushright}
\small UUITP-29/11\\
\small ITP-UH-10/11\\
\small IPhT-T 11/207\\

\date \\
\normalsize
\end{flushright}
\vspace{0.4cm}
\begin{center}

{\LARGE \bf{(Anti-)Brane backreaction beyond \\ \vspace{0.2cm} perturbation theory}} \\

\vspace{1.6 cm} {\large  Johan Bl{\aa}b\"ack$^\dagger$, Ulf
H.~Danielsson$^\dagger$, Daniel Junghans$^\ddagger$, \\
\vspace{0.2cm}Thomas Van Riet$^\flat$, Timm Wrase$^\sharp$ and Marco
Zagermann$^\ddagger$}\footnote{johan.blaback @ physics.uu.se,
ulf.danielsson @ physics.uu.se, daniel.junghans @
itp.uni-hannover.de, thomas.van-riet @ cea.fr, timm.wrase @ cornell.edu, marco.zagermann @ itp.uni-hannover.de }\\

\vspace{1.4 cm}
{${}^\dagger$ Institutionen f{\"o}r fysik och astronomi\\
Uppsala Universitet, Box 803, SE-751 08 Uppsala, Sweden}\\

\vspace{.15 cm}  {${}^\ddagger$ Institut f{\"u}r Theoretische Physik \&\\
Center for Quantum Engineering and Spacetime Research\\
Leibniz Universit{\"a}t Hannover, Appelstra{\ss}e 2, 30167
Hannover, Germany}\\


\vspace{.15 cm} {$\flat$ Institut de Physique Th\'eorique, CEA
Saclay, CNRS URA 2306 \\ F-91191 Gif-sur-Yvette, France}

\vspace{.15 cm}  {${}^\sharp$ Department of Physics, Cornell University, Ithaca, NY 14853, USA}\\



\vspace{1.2cm}

{\bf Abstract}
\end{center}

\begin{quotation}

\noindent We improve on the understanding of the backreaction of
anti-D6-branes in a flux background that is mutually BPS with D6-branes.
This setup is analogous to the study of the
backreaction of anti-D3-branes inserted in the KS throat, but does
\emph{not} require us to smear the anti-branes or do a perturbative
analysis around the BPS background. We solve the full equations of
motion near the anti-D6-branes and show that only two boundary
conditions are consistent with the equations of motion. Upon
invoking a topological argument we eliminate the boundary condition
with regular $H$ flux since it cannot lead to a solution that
approaches the right kind of flux away from the anti-D6-branes. This
leaves us with a boundary condition which has singular, but
integrable, $H$ flux energy density.
\end{quotation}

\newpage

\tableofcontents

\section{Introduction}

Extended objects such as D-branes and orientifold planes play a
crucial role in the search for string theory models that are interesting for
particle physics and cosmology. Generically, these extended objects do not fill out the entire spacetime
but are concentrated on lower-dimensional submanifolds.
In the supergravity limit, this is
described in terms of $\delta$-function-localised source terms that depend on the coordinates
transverse to the brane.

This localised nature of brane-like objects creates a tremendous
complication for solving the equations of motion, because it implies, e.g.,
nontrivial  warp factors, non-constant RR fields and generically
a varying dilaton. For this reason, a simplification
known as smearing is often utilised, where one replaces the
$\delta$-function in the source terms with a smooth, often constant, function that integrates to the same value.
The brane is then evenly distributed along all or some of the
transverse directions. As a result, functions such as the warp factor will have regular
or even constant profiles, since there is no $\delta$-function source anymore that could lead to
singular behaviour.

There are, however, many unanswered questions about the validity of this smearing
procedure. In particular it is unclear what the differences between a
smeared and the corresponding localised solution are. It is also not even
clear under which conditions a smeared solution has a
corresponding localised solution at all. In \cite{Blaback:2010sj} we
initiated a study of these questions\footnote{For related work in a slightly
different context we refer to \cite{Burgess:2011mt, Burgess:2011rv}.}
focusing on sources that are mutually BPS with the background. A typical phenomenon of
BPS solutions is the so-called `no-force' condition, which means that various
ingredients, that are mutually BPS, can be put together without
creating a force on each other. In our BPS examples we found that
the vacuum expectation values (vevs) of the moduli and the cosmological constant of the smeared and localised solutions are the same \cite{Blaback:2010sj}. However, for a non-BPS-setup one can easily imagine that the force on a brane only vanishes when the brane is smeared out. In that case the smearing process might
lead to a cancellation of forces that would otherwise not allow a static solution when the branes are localised.

A potential example of this was discussed in \cite{Blaback:2010sj,
Blaback:2011nz}, inspired by \cite{Douglas:2010rt}. There we constructed (see also \cite{Silverstein:2004id})
a class of smeared solutions where anti-D$p$-branes are
surrounded by fluxes that are not mutually BPS with these anti-D$p$-branes (but rather with D$p$-branes). The particular solution we study here is for the
case of $p=6$ with $F_0$ and $H$ flux. This setup is simple enough
to study the equations of motion for the fully localised anti-D6-branes,
since we can reduce the equations of motion to a set of coupled
ordinary differential equations.

In \cite{Blaback:2011nz} we developed a strong constraint on the
possible boundary conditions for such a solution. The
constraint is topological in nature and restricts the ways the flux type can
change when one moves away from the brane. We demonstrated that the only boundary conditions that
could survive this topological constraint seem to have singularities in fields that are not directly sourced by the anti-D6-branes. But it was not clear whether such boundary conditions exist at all, and, as the argument in
\cite{Blaback:2011nz} was formulated in terms of combinations of several fields, one needs a detailed computation to confirm that such
boundary conditions truly have singularities in fields that do not directly couple to the $\delta$-function sources. This is the scope of the present paper, where we locally compute all possible supergravity solutions near the anti-branes and find exactly one solution that is not ruled out by the topological constraint of \cite{Blaback:2011nz}. This solution has infinite, but integrable, $H$ flux
density, $\e^{-\phi}H^2$ (in Einstein frame).

Our method relies upon expanding the fields in the radial coordinate around the source
and solving the equations of motion order by order. We find
that there are a priori two possible local boundary conditions for extremal
brane sources. The first one is the usual boundary condition with
fluxes around the anti-branes that are mutually BPS with the
anti-branes. In this solution only fields that couple directly to the anti-D6-branes have singularities. This boundary condition is however excluded by our
topological argument. The second boundary condition allows exactly
for the type of singular $H$ flux that evades our topological constraint.

Our results show that singularities in fields that are not directly sourced
by brane-like objects may occur at the level of the non-linear supergravity
equations, and are in general not an artefact of a perturbative analysis around a BPS background
or a partial smearing of the sources. We discuss possible physical interpretations of this
singular behaviour.

This paper is organised as follows. In section 2 we briefly review
the results from our previous paper \cite{Blaback:2011nz}, including
the topological argument that rules out many boundary conditions.
In section 3, we then present the method and the results of our explicit computation of possible boundary behaviours.
The results are then summarized and  discussed  in section 4, where we also
comment on possible implications for the anti-D3-brane backreaction problem
of \cite{Kachru:2002gs, DeWolfe:2008zy,
McGuirk:2009xx, Bena:2009xk, Dymarsky:2011pm, Bena:2011hz,
Bena:2011wh}. The appendices contain our conventions and the details of our calculations.

\section{Review of previous results}

In order to establish our notation and conventions we start out by reviewing the solutions with smeared anti-D6-branes. Then we present the most general ansatz for the fields in the localised case and derive a topological constraint that must be satisfied by all solutions. We refer to appendix \ref{sec_smeared} and \cite{Blaback:2011nz} for more details.

\subsection{The smeared solution}

For the smeared solution, we take the 10D spacetime to be the direct product $AdS_7 \times S^3$ and add anti-D6-branes that fill $AdS_7$ and are uniformly smeared along the $S^3$. To cancel their charge we turn on non-zero $F_0$ and $H$ flux. Choosing the $AdS_7$ radius such that the $AdS_7$ Ricci scalar is $-42$ and the radius of the 3-sphere such that the $S^3$ has Ricci scalar $6$, one finds that all equations of motion are satisfied, if we choose
\begin{equation}
H=h \star_3 1\,,\qquad F_0 = \frac{2}{5} \e^{-\frac{7}{4}\phi_0} h\,,
\end{equation}
where $h$ is a constant and $\phi_0$ is the constant value of the dilaton. The tadpole condition implies that
the charge of the anti-D6-branes is given by $Q =-N_{D6} \mu_6 = -\frac{2}{5} h^2 \e^{-\frac{7}{4}\phi_0}$.

In \cite{Blaback:2011nz} it was shown that all closed string moduli that are left-invariant under the action of $\SU(2)$ (regarding $S^3$ as Lie group $\SU(2)$) have positive semi-definite masses.

\subsection{The ansatz for the localised solution}

In order to study the case of localised anti-D6-branes, we assume that all branes are localised at one or both of the poles of the $S^3$. This allows us to make a highly symmetric ansatz that preserves an $\SO(3)$ rotational symmetry which reduces the
equations of motion to ODE's. The most general metric ansatz is
\begin{equation}\label{metric}
\d s^2_{10} = \e^{2A(\theta)} \d s_{AdS_7}^2 + \e^{2B(\theta)}\left(\d \theta^2 + \sin^2(\theta)
\d \Omega^2\right).
\end{equation}
The corresponding flux ansatz is
\begin{align}
& H=\lambda F_0 \e^{\tfrac{7}{4}\phi}\star_3 1\,, \label{eq:Hflux}\\
& F_2 = \e^{-\tfrac{3}{2}\phi-7A}\star_3\d\alpha\,,
\end{align}
where $\phi, \lambda$  and $\alpha$ are now functions depending on
$\theta$, $\star_3$ contains the conformal factor and we take $F_0$ to be constant. This is the most general
ansatz compatible with the form of the equations of motion and our
symmetries. Hence, the problem is reduced to finding a set of five
unknown functions $A, B, \phi, \lambda, \alpha$ depending on
$\theta$ and obeying coupled second-order differential equations,
which we now derive.

The $F_2$-Bianchi identity reads
\begin{equation}
\frac{\Bigl(\e^{-\tfrac{3}{2}\phi-7A+B}\sin^2\theta
\alpha'\Bigr)'}{\e^{3B}\sin^2\theta}=\e^{\tfrac{7}{4}\phi}\lambda
F_0^2 +Q\delta(D6)\,, \label{eom1}
\end{equation}
where a prime $'$ denotes the derivative with respect to $\theta$,
e.g., $A'=\d A/\d \theta$. The $F_2$ equation of motion $\d (\e^{\frac{3}{2}\phi} \star_{10} F_2)=0$ is automatically satisfied. The $H$ equation of motion, $\d(\e^{-\phi} \star_{10} H) = - \e^{\frac{3}{2}\phi} F_0 \star_{10} F_2$, allows us to
eliminate $\alpha$ in terms of $\lambda$
\begin{equation}\label{alphaversuslambda}
\alpha=\e^{\tfrac{3}{4}\phi +7A}\lambda \,,
\end{equation}
where we have used that through a shift in $\alpha$ we can always set the integration constant to zero.
The dilaton and Einstein equations give the following differential
equations
 \allowdisplaybreaks
\begin{align}
& \frac{\Bigl(\e^{7A+B}\sin^2\theta
\phi'\Bigr)'}{\e^{7A+3B}\sin^2\theta}=\e^{\tfrac{5}{2}\phi} F_0^2
\bigl(\tfrac{5}{4}-\tfrac{\lambda^2}{2} \bigr) + \tfrac{3}{4}\e^{-
14A - 2B - \tfrac{3}{2}\phi}(\alpha')^2 +
\tfrac{3}{4}\e^{\tfrac{3}{4}\phi} T\delta(D6)\,, \label{eom2}\\
&  -96\e^{-2A}  -16 \e^{-2B} \lp 7(A')^2 + A'B' +
\frac{(\sin^2\theta A')'}{\sin^2\theta} \rp = \,
\e^{\tfrac{5}{2}\phi} F_0^2 (1-2\lambda^2) \nonumber\\&-
\e^{-14A-2B-\tfrac{3}{2}\phi}(\alpha')^2  -
\e^{\tfrac{3}{4}\phi}T\delta(D6)\,, \label{eom3}\\
&  2  - \frac{(\sin^2\theta B')'}{\sin^2\theta} - 7(A')^2 - B''
-7A'' + 7A'B' =  \tfrac{1}{2}(\phi')^2 +
\tfrac{1}{16}\e^{\tfrac{5}{2}\phi + 2
B}\,F_0^2\Bigl(1+6\lambda^2\Bigr)\nonumber\\ & -
\tfrac{1}{16}\e^{-14A-\tfrac{3}{2}\phi}(\alpha')^2
+\tfrac{7}{16}\e^{\tfrac{3}{4}\phi +2B}T\delta(D6)\,, \label{eom4}\\
& 2  - (B')^2 -\frac{(\sin^2\theta B')'}{\sin^2\theta}
-\cot\theta(B+7A)' -7 A'B' = \tfrac{1}{16}\e^{\tfrac{5}{2}\phi + 2
B}\,F_0^2\Bigl(1+6\lambda^2\Bigr) +  \nonumber\\
& \tfrac{7}{16}\e^{-14A-\tfrac{3}{2}\phi}(\alpha')^2 +
\tfrac{7}{16}\e^{\tfrac{3}{4}\phi +2B}T\delta(D6)\,. \label{eom5}
\end{align}
Since the external and internal metric have an undetermined warp/conformal factor in front,
we can normalise the $AdS_7$ and $S^3$ scales to our liking. To be
conform with the previous paper \cite{Blaback:2011nz} we have taken the
$AdS_7$ Ricci scalar to be $-42$ and the $S^3$ Ricci scalar to be $6$.

\subsection{A topological constraint}\label{topological}
At any point away from the sources we can combine
\eqref{eom1} and \eqref{alphaversuslambda} to find
\begin{equation}
\frac{\Bigl(\e^{-\tfrac{3}{2}\phi-7A+B}\sin^2\theta
\Bigr)'}{\e^{3B}\sin^2\theta}\alpha' + \e^{-\tfrac{3}{2}\phi-7A -2B}
\alpha''=\alpha \e^{\phi-7A} F_0^2\,.
\end{equation}
This equation tells us that whenever $\alpha'=0$, the sign of $\alpha$ determines whether the extremum is a maximum or minimum. Explicitly we have that at any extremum away from a source
\be\label{eq:topological}
\sgn \alpha'' = \sgn \alpha.
\ee
Whenever $H \propto \star_{9-p} F_{6-p}$, a similar constraint for anti-D$p$-branes with $p<6$ can be derived in other dimensions and other geometries, including cone-like non-compact geometries.

\section{Boundary conditions}

In \cite{Blaback:2011nz}, we investigated to what extent a smeared
solution of a $\overline{D6}$-brane on $AdS_7\times S^3$ can
correspond to a truly localised solution. The simple topological
argument discussed in subsection \ref{topological} rules out
solutions for many given boundary conditions of the 10D supergravity
fields near the $\overline{D6}$-brane, including the standard
BPS-type boundary behaviour. For certain singular, non-standard,
boundary conditions, however, the topological argument can be
circumvented. It is the purpose of this section to determine all the
possible boundary conditions near the $\overline{D6}$-branes and to
revisit the discussion of \cite{Blaback:2011nz}.

To this end we will momentarily put aside all intuition or
preconceptions about what the right boundary conditions should be
and let the 10D supergravity equations tell us what all the
potential boundary behaviours are. The result will turn out
surprisingly simple and will be further discussed in section
\ref{results}.

To uncover the possible near brane behaviours of the 10D
supergravity fields in our setup, we expand the vacuum equations
around $\theta=0$, i.e., outside of, but close to, the source, with
the following general ansatz:
\begin{eqnarray}
& \e^{-A(\theta)} = a_0 \theta^A + a_1 \theta^{A+\zeta} + a_2
\theta^{A+\xi} + \ldots , \qquad \e^{-2B(\theta)} = b_0 \theta^B +
b_1 \theta^{B+\zeta} + b_2 \theta^{B+\xi} + \ldots, & \notag \\ &
\e^{-\frac{1}{4} \phi(\theta)} = f_0 \theta^F + f_1 \theta^{F+\zeta}
+ f_2 \theta^{F+\xi} + \ldots, \qquad \lambda(\theta) = \lambda_0
\theta^L + \lambda_1 \theta^{L+\zeta} + \lambda_2 \theta^{L+\xi} +
\ldots \label{eq:aaa}
\end{eqnarray}
Here, $A,B,F,L$ and $\zeta < \xi < \ldots$ are unknown (possibly not
integral) real numbers. In this expansion we assume that $\e^{-A}$,
$ \e^{-\frac{1}{4}\phi} $, $\e^{-2B} $, $\lambda$ do not have
essential singularities, i.e., that there is really a finite leading
power of $\theta$ for each function.

The coefficients $a_0, b_0, f_0, \lambda_0$ are taken to be non-zero
such that $A,B,F,L$ by definition determine the leading order
divergences of the fields. Since they correspond to the expansion of
exponential functions, $a_0, b_0, f_0$ must in addition be
non-negative. However, we allow any of the sub-leading order
coefficients (such as e.g. $a_1,b_5,\lambda_2,$ etc.) to be possibly
zero, so that the steps between the different powers of $\theta$ in
the various expansion series need not be the same at all orders and
for all functions. \footnote{For example, if $\zeta=\frac{1}{2},
\xi=1$ and $a_1\neq 0, f_1=0, f_2\neq 0$, we would have
$e^{-A(\theta)}=a_0 \theta^A +a_1\theta^{A+\frac{1}{2}}+\ldots$ and
$e^{-\frac{1}{4}\phi(\theta)}=f_0\theta^F+f_2\theta^{F+1}+\ldots$}

Plugging this general ansatz into the equations of motion
\eqref{eom1} to \eqref{eom5}, we can now explicitly check which
choices for the powers $A,B,F,L,\zeta,\xi,\ldots$ and the
coefficients $a_n,b_n,f_n,\lambda_n$ are consistent. Although this
computation is rather lengthy (cf. App. \ref{app_bc}), the result is
surprisingly simple: we find that possible boundary conditions are
locally restricted to only five different cases. As is detailed in
section \ref{sec:sourcedet}, a careful analysis of the field
behaviour near $\theta=0$ yields the tension and the RR-charge of
the $\delta$-type source that must be present at the origin in order to
support the solution across the pole. It turns out that one of the
five local solutions leads to contradictory results for the tension and another one is singular at the pole without the presence of
local sources there. We therefore discard these cases and only
list the remaining three possibilities, which decompose into two
familiar ones with straightforward interpretation and one novel boundary condition:
\begin{itemize}
\item The smooth solution with no sources sitting at the pole:
\begin{equation}
L=A=B=F=0.
\end{equation}
\item  The standard BPS boundary condition
 \begin{equation}
L=0, \qquad A = -\tfrac{1}{16}, \qquad B = \tfrac{7}{8}, \qquad F = -\tfrac{3}{16}, \qquad \lambda_0 = \pm 1  \label{bc2}
\end{equation}
supported by extremal branes with $|Q|=T$.
\item A previously unknown boundary condition with divergent $\lambda(\theta)$
\begin{equation}
L=-1, \qquad A = -\tfrac{1}{16}, \qquad B = \tfrac{7}{8}, \qquad F = -\tfrac{3}{16}, \label{bc3}
\end{equation}
which is again supported by extremal branes with $|Q|=T$.
\end{itemize}

The three physical boundary conditions\footnote{Since the last two solutions we have found are divergent, the supergravity approximation inevitably breaks down near $\theta=0$. The requirement that the low energy effective action is applicable for a certain range of $\theta$ imposes constraints on the parameters in the expansion \eqref{eq:aaa}. String loop corrections can be suppressed for sufficiently large $f_i$ while curvature corrections are small for sufficiently small $a_i$ and $b_i$. Increasing the number of anti-branes simultaneously makes the string coupling and the curvature smaller.} are summarized in table
\ref{table1} and derived in detail in App. \ref{app_bc}.
\footnote{Note that our general ansatz can even be extended to also
allowing a logarithmic behaviour of the fields that does not sum up
to some power of $\theta$ according to $\theta^k = e^{k\ln\theta}$ ,
e.g. $ \e^{-A(\theta)} = a_0 \theta^A (\ln \theta)^{\tilde A} +
\ldots, \e^{-2B(\theta)} = b_0 \theta^B (\ln \theta)^{\tilde B} +
\ldots $ with generic coefficients and accordingly for the other
fields. Allowing such logarithms, one finds that all powers of
logarithms have to be zero in the leading order terms of the fields
so that no further boundary conditions arise other than those listed
above. The computation is analogous to the one shown in App.
\ref{app_bc}.}

\begin{table}[ht]
\begin{center}
  \begin{tabular}{ |c | c | c | c || c | c || c | c || c | }
    \hline
    $L$ & $A$ & $B$ & $F$ & $\lambda(0)$ & $\alpha$ & source & $\frac{|Q|}{T}$ & valid \\ \hline\hline
    $0$ & $0$ & $0$ & $0$ & ? & $ \mathcal{O}(1) $ & none & - & \checkmark \\ \hline
    $0$ & $-\frac{1}{16}$ & $\frac{7}{8}$ & $-\frac{3}{16}$ & $\pm 1$ & $ \mathcal{O}(\theta^1) $ & $D/\bar D$ & $1 $ & only locally \\ \hline
    $-1$ & $-\frac{1}{16}$ & $\frac{7}{8}$ & $-\frac{3}{16}$ & $\pm \infty $ & $ \mathcal{O}(1) $ & $D/\bar D$ & $1$  & \checkmark \\ \hline
    \end{tabular}
\caption{The different boundary conditions that are locally allowed by the equations of motion. The first row corresponds to the situation without a source at the pole in question. This can be part of a global solution only when there are sources at the other pole so as to cancel the global tadpole. The usual BPS boundary condition (second row) is excluded globally due to the topological constraint of \cite{Blaback:2011nz} (see below). The third row shows the only boundary condition with an extremal source at the pole that also evades the topological contraint. If a global solution exists, it thus has to approach a solution of the third type for at least one of the poles.}
\label{table1}
\end{center}
\end{table}

Let us now discuss the newly found boundary condition \eqref{bc3}.
Using the scaling of the fields near the pole, one finds that the
behaviour of the function $\alpha$ (see eq.
(\ref{alphaversuslambda})) is
\begin{equation}
\alpha = \alpha_0 + \alpha_1 \theta + \ldots = \frac{\lambda_0}{a_0^7f_0^3} \pm \frac{1}{a_0^7f_0^3} \theta + \mathcal{O}(\theta^2).
\end{equation}
As $\lambda_0,a_0,f_0$ are by definition all non-zero,
$\alpha(\theta)$ is finite and non-zero near the pole. The sign of
$\lambda_0$, however, is not determined by the equations of motion,
so that $\alpha(\theta)$ may start out positive or negative at the
pole, depending on the chosen sign of $\lambda_{0}$.

\begin{figure}[h!]
\centering
\includegraphics[width=1\textwidth]{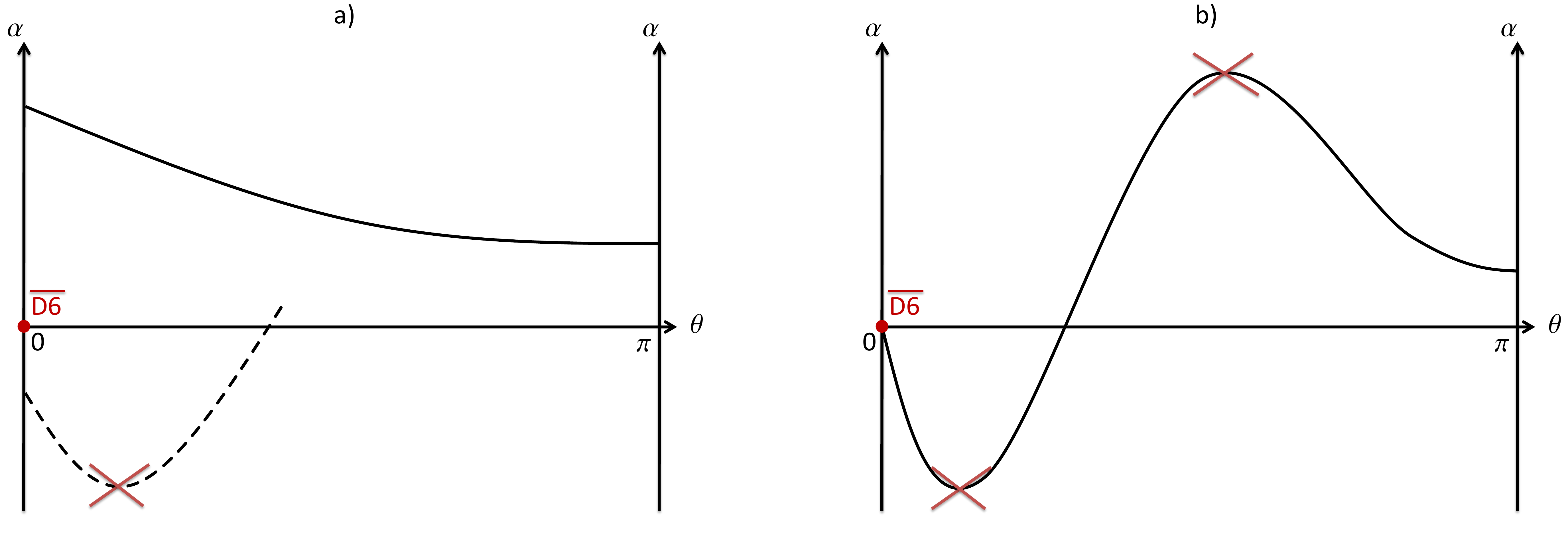}
\caption{a) This plot shows the boundary conditions (\ref{bc3}) that
evade the constraint \eqref{eq:topological} provided that $\alpha(0)>0$ (solid line). For $\alpha(0)<0$ (dashed line), boundary condition (\ref{bc3}) would not work, because $\alpha$ would have to violate \eqref{eq:topological} in order to become positive to cancel the global $\overline{D6}$ tadpole. b) This plot shows that the usual BPS boundary conditions near the
$\overline{D6}$ are likewise excluded due to the forbidden extrema (crosses) that would be necessary to ensure overall tadpole cancellation.\label{boundary}}
\end{figure}

The sign of $\alpha_1=\alpha^{\prime}(0)$, on the other hand, cannot
be chosen freely but is determined by the sign of the charge of the
source that sits at $\theta=0$. As shown in the appendix, the charge
and the tension of sources that are compatible with boundary
condition \eqref{bc3} are given by
\begin{equation}
Q = \frac{a_0^7 f_0^6}{\sqrt{b_0}} \alpha_1 = \pm \frac{f_0^3}{\sqrt{b_0}}, \qquad T = \frac{f_0^3}{\sqrt{b_0}}.
\end{equation}
The sign of $\alpha^\prime$ near the pole is hence negative for
$Q<0$, i.e. for $\overline{D6}$-branes (and positive for
D6-branes with charge $Q>0$). As is shown in figure \ref{boundary} a) for the case of $\overline{D6}$-branes,
this boundary condition thus evades the topological constraint only
for positive $\alpha_0$. For negative $\alpha_0$ (dashed line) $\alpha$ cannot become positive without having a minimum that is forbidden by the topological constraint \eqref{eq:topological}. Thus $\alpha$ and therefore by \eqref{alphaversuslambda} $\lambda$ are always negative. Since $Q$ is also negative Gauss law is violated, since the left-hand side of \eqref{eom1} integrated over the $S^3$ gives zero while the right-hand side is negative.

It is instructive to compare the above boundary condition
(\ref{bc3}) with the standard BPS-behaviour (\ref{bc2}). The main
difference is the leading power, $L$, of $\theta$ in
$\lambda(\theta)$, which is $L=0$ in the BPS case and $L=-1$ in
(\ref{bc3}). Unlike the case (\ref{bc3}) where $\alpha(\theta)$ must have a
non-vanishing constant term, the BPS-case starts instead with the linear
power in $\theta$, c.f. (\ref{alphabehavior2}), whose sign is again
determined by the sign of the charge $Q$ of the source, cf.
(\ref{chargelambda2}). For a $\overline{D6}$-brane, this sign is
again negative with the aforementioned consequences for the
non-existence of a global solution, see figure \ref{boundary} b).

\section{Summary and interpretation of results}\label{results}
In the previous section we have shown that there is only one local
boundary behaviour that could evade the constraint of
\cite{Blaback:2011nz} which is reviewed in subsection \ref{topological}. This boundary condition (cf. row 3 of table
\ref{table1}) leads to \be e^{-\phi} H^2 \propto \theta^{-\frac18},
\ee i.e. a $H$ flux density that diverges at the source. Note
however that the action is finite since \be \sqrt{g_{10}} e^{-\phi}
H^2 \propto e^{7A+3B} \sin^2(\theta) \theta^{-\frac18} \propto
\theta. \ee
Our original question as to whether the smeared solution has
a sensible localised counterpart, can now be rephrased
in terms of two much more precise questions:\\
(i) Do the singular boundary conditions (\ref{bc3}) that evade the topological constraint
really integrate to a reasonable global solution of the supergravity equations?\\
(ii) If yes, how similar is this localised solution compared to the smeared one, e.g. regarding
moduli vevs or the cosmological constant?\\
(iii) Supposing it is part of a full solution, what is the physical interpretation of the singular behaviour of the
$H$ flux density at the anti-brane location?

The first question is a difficult numerical problem that is exacerbated by
the fact that the boundary conditions at the pole still leave five undetermined parameters.
Our preliminary searches for numerical solutions so far only led to singularities
in the bulk of the three-sphere, but they also covered a far too limited parameter range to be
in any way conclusive, which also makes quantitative statements about question (ii) difficult.

We will therefore focus, in the remainder of this paper, on the third question and discuss
possible interpretations of the singular boundary behaviour of a hypothetical global solution.
Before we do this, however, we would like to put our results in perspective and discuss
the relation to a few similar setups of recent interest.

While our determination of the boundary conditions in the previous section  was for the manifold $AdS_7\times S^3$,
we have checked that one obtains the same result if one
sets the curvature of the $AdS_7$ space and the $S^3$ to zero.
This
means that fully backreacted anti-D6-branes in flat space with $H$ flux along the three transverse directions and in the presence of
$F_0$ lead to the same singularity in the $H$ flux density (if one
demands that the flux far away from the anti-D6-branes becomes
mutually BPS with D6-branes). At first sight this non-compact case might seem to
make contact with the fractional brane solutions of \cite{Herzog:2000rz} for the case $p=6$.
This is not a valid interpretation, though, as the singularities in a corresponding D6-brane background cannot be
deformed away as e.g. for the conifold in the D3-brane case. This means one would always have the D6-branes sitting at the origin
so that adding an anti-D6-brane there would not give a stable setup to begin with. \footnote{The  non-compact BPS background with D6-branes has some problematic features. Apart from the fact that the singularity at small radius cannot be resolved (contrary to the KS background), there is also a naked singularity at large radius. This requires a cut-off at large radius. Such a cut-off could be naturally provided in a compact setting.}

The analogous fractional brane backgrounds for some other values of $p$, however, do not
suffer from this problem and can be smoothed off at the tip before brane singularities are reached.
Putting an anti-brane at the tip of such a pure flux background, one could therefore ask
what kind of boundary behaviour would be allowed by the supergravity equations, thereby addressing the recent works on anti-D3-branes in the
Klebanov-Strassler (KS) throat \cite{Kachru:2002gs, DeWolfe:2008zy,
McGuirk:2009xx, Bena:2009xk, Dymarsky:2011pm, Bena:2011hz,
Bena:2011wh} or to anti-M2-branes on warped Stenzel spaces
\cite{Bena:2010gs, Massai:2011vi} or to anti-D2-branes on the CGLP
background \cite{Giecold:2011gw}. The backreaction of the
anti-branes in all these cases has been intensively studied to first
order in $p/M$, where $p$ is the number of anti-branes and $M$ the
background flux. The solutions have singularities in fields that do
not couple directly to the anti-branes. An important question is
whether such singularities are still present in the case of fully
backreacted anti-branes or whether they are artefacts of approximation schemes.  Although recently new tools have been
developed that make it possible to calculate
higher order corrections \cite{Gandhi:2011id}, a fully backreacted solution for
anti-D3-branes in the KS throat seems clearly out of reach.

Interestingly, even though our solutions in the non-compact case cannot be directly interpreted in terms of a fractional brane background, they are yet very much analogous to the above backgrounds from the pure supergravity point of view.
In this analogous setup we answered the above question and have shown that
singularities that are not directly sourced by the anti-branes have
to be present in the fully backreacted solution. We take our results as indication that
the same is true for the other backreacted anti-brane solutions as well.
The singular flux found in \cite{Kachru:2002gs, DeWolfe:2008zy,
McGuirk:2009xx, Bena:2009xk, Dymarsky:2011pm, Bena:2011hz,
Bena:2011wh} should therefore not be an artefact of the $p/M$-expansion or of the partial smearing
of the anti-branes used in those works, but should be taken seriously as a feature of the non-linear supergravity solutions.

We now come to the physical interpretation of the singularities themselves. It is
instructive to consider the above solutions in terms of forces and
their balance:

\subsubsection*{A `no no force' problem?}
The flux combination $H \wedge F_{6-p}$ appears in the $F_{8-p}$ Bianchi identity exactly like smeared D$p$-branes or
anti-D$p$-branes, depending on the nature of the flux (i.e.~ISD or IASD for $p=3$). This leads to complications whenever we add anti-branes to a background that is mutually BPS with branes. Let us for clarity discuss this for the case of
anti-D3-branes in the KS throat solution. The KS throat is defined
by ISD flux. If one inserts a D3-brane in this background then a
backreacted solution is relatively simple because the D3-brane
feels no force. The charge of the ISD flux repels the brane but at the same time
there is a gravitational attraction between the flux and the brane
that exactly cancels this. This is a typical no-force condition for
extremal/BPS solutions in supergravity. However, upon inserting an
anti-brane the delicate balance is lost. The ISD flux is both
attracted electromagnetically and gravitationally to the anti-brane.

A possible way out would be for the flux to change to IASD type
around the anti-brane. But then one would expect perturbative
instabilities in the flux cloud as it has to smoothly interpolate
from IASD to ISD type. This is quantified by the topological
constraint we have found in \cite{Blaback:2011nz}.

Another way out is given by the force related to the gradient energy
of the fluxes. For a given flux number the energetically favoured
configuration is a smooth flux distribution. The more gradient there
is in the flux profile the more energy this costs. Hence, this
provides a natural counter-force that acts against the accumulation of the flux near the
anti-brane due to the electromagnetical and gravitational attraction.

Considering the flux profile of our solution we notice that there is
a piling up of ISD flux\footnote{ISD refers to the $F_3$ and $H$
flux but we equally use this for the general $F_{6-p}$ and $H$ flux.
ISD then means flux that carries D$p$ charge.} near the anti-brane (cf. figure \ref{boundary} a)).
This is consistent with our physical picture of the anti-brane attracting the flux cloud. Note, however, that we find an infinite, but integrable, flux energy density $e^{-\phi} H^2$. Singularities that are not directly sourced by some $\delta$-function in the action often indicate that the setup has an instability. One could interpret such an instability in our setup in the following two ways:

\subsubsection*{Interpretation 1: Brane nucleation}
The gradient energy does provide the required counteracting force
and the static configuration is given by our solution. The $H$ flux
singularity might then get resolved in string theory by the open
string degrees of freedom as we now explain: Our setup is
perturbatively unstable in the open string sector due to the Myers
effect \cite{Myers:1999ps} in which D-branes can undergo a
transition and change into a higher dimensional brane that wraps a
trivial cycle.\footnote{This instability is not an issue in the case of only one brane, but this case
would also correspond to small volume and hence raise doubts on the validity of the supergravity approximation we are using here.} For example in \cite{Kachru:2002gs} it was shown that
the $p$ anti-D3-branes at the tip of the KS throat relax into an
NS5-brane. It is conceivable that such a transition cures the
remaining unphysical singularities \cite{Dymarsky:2011pm}. Again due
to the complexity of the KS geometry it seem very hard to solve the
fully backreacted equations of motion after such a transition.
However, it should be possible to study the analogue of such a
transition in our simple setup. The puffed-up background is probably unstable itself
since, in contrast with the KS solution, we have not been able to
smoothen out the brane singularities of the BPS solution. Therefore
we had to take as sources at the origin pure anti-branes. This
implies that we are far away from the BPS solution, and in analogy
with \cite{Kachru:2002gs} this implies that the puffed-up system is
perturbatively unstable against brane-flux annihilation. If true
then the puffed-up solution might again have singularities.

\subsubsection*{Interpretation 2: No balance of forces}
Another explanation could be that the gradient energy in the fluxes is not able to counteract
the attraction of the flux towards the anti-brane and this is the
reason that there are no solutions with finite $H$ flux density. The true solution would then be time-dependent with regular $H$ profile at finite times. It is known that
perturbative unstable systems, when forced into a static ansatz,
develop unwanted singularities \cite{Dvali:2002pe}, and it seems possible
to interpret our singularity in this way. If this was correct, this
would imply that the construction of dS vacua in string theory through
uplifting by anti-D3-branes \cite{Kachru:2003aw, Balasubramanian:2005zx} would be
inconsistent. This would be the most
exciting outcome and could point to string theory being
very constraining when it comes to constructing meta-stable de
Sitter vacua (which seems to be the case for the construction of dS solutions at
the classical level see e.g. \cite{Caviezel:2009tu,Danielsson:2011au} and references therein for recent discussions).

Both interpretations have in common that the singularity in the flux
is attributed to an instability. In the second (more controversial)
interpretation, the instability is already present in the degrees of
freedom captured by the bulk supergravity action. In the first
interpretation, the instability is only visible upon inclusion of
the open string degrees of freedom.

\subsection*{Acknowledgements}
We benefitted from useful discussions and emails with I. Bena, A.
Dymarsky, G. Dvali, G. Giecold, M. Gra\~na, J. Maldacena, S. Massai, L.
McAllister, F. Orsi, A. Puhm, G. Shiu and B. Vercnocke. The work of
J.B. is supported by the G{\"o}ran Gustafsson Foundation. U.D. is supported by the Swedish Research
Council (VR) and the G\"oran Gustafsson Foundation. D.J. and M.Z.
are supported by the German Research Foundation (DFG) within
the Cluster of
Excellence ``QUEST''. The work of
T.V.R. is partially supported by the ERC Starting Independent Researcher Grant
259133 -ObservableString. T.W. is supported by a Research Fellowship (Grant number WR 166/1-1) of the German Research Foundation (DFG), by the Alfred P. Sloan Foundation and by the NSF under grant PHY-0757868.

\appendix
\section{Type IIA supergravity with $D6/\overline{
D6}$-branes} \label{sec_smeared}

We use the same conventions as in \cite{Blaback:2011nz}.  Throughout
the paper $a,b$ are 10D indices, $\mu, \nu$ are external and  $i,j$
are internal indices. We present the equations in Einstein frame and
have put $F_4=0$. The trace reversed Einstein equation is then
\begin{align}
\label{eoms_einstein} R_{ab} = & \tfrac{1}{2} \partial_a \phi
\partial_b \phi + \tfrac{1}{2} \e^{-\phi} |H|^2_{ab}-\tfrac{1}{8}
\e^{-\phi} g_{ab} |H|^2 + \tfrac{1}{16} \e^{\tfrac{5}{2} \phi}
g_{ab} F_0^2 \\ \notag &{} + \tfrac{1}{2} \e^{\tfrac{3}{2} \phi}
|F_2|^2_{ab} -\tfrac{1}{16} \e^{\tfrac{3}{2} \phi} g_{ab} |F_2|^2
+\tfrac{1}{2}(T^{loc}_{ab} - \tfrac{1}{8}g_{ab}T^{loc}),
\end{align}
where $|A|^2_{ab} \equiv \tfrac{1}{(p-1)!}\,A_{a a_2\ldots
a_p}A_{b}^{\,\,a_2\ldots a_p}$, $|A|^2 \equiv
\tfrac{1}{p!}\,A_{a_1\ldots a_p}A^{a_1\ldots a_p}$.

The non-vanishing part of the local stress tensor is given by
\begin{equation}
T_{\mu\nu}^{loc} = - \e^{\tfrac{3}{4}\phi} \mu_6 g_{\mu\nu}
\delta(D6),
\end{equation}
where $\mu_6$ is a positive number, and $\delta(D6)$ is the $\delta$-function with support on the $D6$-branes world volume(s). The
dilaton equation of motion is given by
\begin{equation}
\label{eoms_dilaton} \nabla^2 \phi= - \tfrac{1}{2} \e^{-\phi} |H|^2
+ \tfrac{5}{4} \e^{\tfrac{5}{2} \phi} F_0^2 +\tfrac{3}{4}
\e^{\tfrac{3}{2} \phi} |F_2|^2 + \tfrac{3}{4} \e^{\tfrac{3}{4}\phi}
\mu_6 \delta(D6).
\end{equation}
The Bianchi identities for the field strengths are
\begin{equation}
\label{eoms_bianchi} \d H = 0, \qquad \d F_0  = 0, \qquad \d F_2 = H
F_0 \pm \mu_{6} \delta_{3},
\end{equation}
where the upper sign of the source term is for $D6$-branes and the
lower sign for anti-$D6$-branes.  $\delta_{3}$ is shorthand for the
normalized volume 3-form transverse to the $D6$-brane multiplied by
$\delta(D6)$, $\delta_3=\delta(D6)\star_3 1$.
\begin{equation}
\int \delta(D6) \star_3 1 = 1.
\end{equation}
The equations of motion for $H$ and $F_2$ read
\begin{equation}
\label{eoms_fluxes} \d(\e^{-\phi}\star H) =
-\e^{\frac{3}{2}\phi}\star F_2 \w F_0, \qquad \d \lp
\e^{\tfrac{3}{2} \phi} \star F_2 \rp = 0.
\end{equation}
In the smeared limit we have
\begin{equation}
\delta(D6) \to \frac{1}{V},
\end{equation}
where $V$ is the volume of the space transverse to the branes.

\section{Determining the boundary conditions}
 \label{app_bc}

We use the following combinations of the equations of motion \eqref{eom1} to \eqref{eom5} which we spell out for convenience ($a(\theta) = \e^{-A(\theta)}, b(\theta) = \e^{-2B(\theta)}, f(\theta) = \e^{-\frac{1}{4} \phi(\theta)}$): a combination of the external Einstein equation \eqref{eom3} and the dilaton equation \eqref{eom2} such that the source term cancels out which yields
\begin{align}
& 4\tfrac{f(\theta)^{\prime\prime}}{f(\theta)} - 4 \tfrac{f(\theta)^{\prime 2}}{f(\theta)^2} + 8\tfrac{\cot(\theta) f(\theta)^\prime}{f(\theta)} - 28 \tfrac{a(\theta)^\prime f(\theta)^\prime}{a(\theta)f(\theta)} - 2\tfrac{b(\theta)^\prime f(\theta)^\prime}{b(\theta)f(\theta)} - 12 \tfrac{a(\theta)^{\prime\prime}}{a(\theta)} + 96 \tfrac{a(\theta)^{\prime 2}}{a(\theta)^2} - 24 \tfrac{\cot(\theta) a(\theta)^\prime}{a(\theta)} \notag \\ &{} + 6\tfrac{a(\theta)^\prime b(\theta)^\prime}{a(\theta)b(\theta)} + 72 \tfrac{a(\theta)^2}{b(\theta)} - 2 \tfrac{\lambda(\theta)^2 F_0^2}{b(\theta)f(\theta)^{10}} + 2 \tfrac{F_0^2}{b(\theta)f(\theta)^{10}} = 0, \label{eq:e1}
\end{align}
the dilaton equation \eqref{eom2}
\begin{align}
& 4\tfrac{f(\theta)^{\prime\prime}}{f(\theta)} - 4 \tfrac{f(\theta)^{\prime 2}}{f(\theta)^2} + 8 \tfrac{\cot(\theta) f(\theta)^\prime}{f(\theta)} - 28 \tfrac{a(\theta)^\prime f(\theta)^\prime }{a(\theta)f(\theta)} - 2 \tfrac{b(\theta)^\prime f(\theta)^\prime}{b(\theta)f(\theta)} - \tfrac{1}{2} \tfrac{\lambda(\theta)^2 F_0^2}{b(\theta)f(\theta)^{10}} + \tfrac{5}{4} \tfrac{F_0^2}{b(\theta)f(\theta)^{10}} + \tfrac{147}{4} \tfrac{\lambda(\theta)^2 a(\theta)^{\prime 2}}{a(\theta)^2} \notag \\ &{} + \tfrac{63}{2} \tfrac{\lambda(\theta)^2 a(\theta)^\prime f(\theta)^\prime}{a(\theta)f(\theta)} - \tfrac{21}{2} \tfrac{\lambda(\theta) a(\theta)^\prime \lambda(\theta)^\prime}{a(\theta)} + \tfrac{27}{4} \tfrac{\lambda(\theta)^2 f(\theta)^{\prime 2}}{f(\theta)^2} - \tfrac{9}{2} \tfrac{\lambda(\theta)f(\theta)^\prime \lambda(\theta)^\prime}{f(\theta)} + \tfrac{3}{4} \lambda(\theta)^{\prime 2} = -\tfrac{3}{4} T \tfrac{\sqrt{b(\theta)} \delta(\theta)}{f(\theta)^3 \sin^2(\theta)},
\label{eq:e2}
\end{align}
the Bianchi identity \eqref{eom1}
\begin{align}
& 7\tfrac{\lambda(\theta)a(\theta)^{\prime\prime}}{a(\theta)} - 7\tfrac{\lambda(\theta)a(\theta)^{\prime 2}}{a(\theta)^2} + 14 \tfrac{\cot(\theta) \lambda(\theta) a(\theta)^\prime}{a(\theta)} + 3\tfrac{\lambda(\theta)f(\theta)^{\prime\prime}}{f(\theta)} + 6 \tfrac{\lambda(\theta)f(\theta)^{\prime 2}}{f(\theta)^2} + 6\tfrac{\cot(\theta) \lambda(\theta) f(\theta)^\prime}{f(\theta)} - \lambda(\theta)^{\prime\prime} \notag \\ &{} - 2 \cot(\theta) \lambda(\theta)^\prime + \tfrac{\lambda(\theta) F_0^2}{b(\theta)f(\theta)^{10}} - \tfrac{7}{2} \tfrac{\lambda(\theta) a(\theta)^\prime b(\theta)^\prime}{a(\theta)b(\theta)} - \tfrac{3}{2} \tfrac{\lambda(\theta) b(\theta)^\prime f(\theta)^\prime}{b(\theta)f(\theta)} + \tfrac{1}{2} \tfrac{b(\theta)^\prime \lambda(\theta)^\prime}{b(\theta)} + 21 \tfrac{\lambda(\theta) a(\theta)^\prime f(\theta)^\prime}{a(\theta)f(\theta)} \notag \\ &{} + 7 \tfrac{a(\theta)^\prime \lambda(\theta)^\prime}{a(\theta)} = - Q \tfrac{\sqrt{b(\theta)} \delta(\theta)}{f(\theta)^3 \sin^2(\theta)},
\label{eq:e3}
\end{align}
a combination of the internal Einstein equations \eqref{eom4} and \eqref{eom5} such that all $F_0^2$-dependent terms cancel out which yields
\begin{align}
&{} - 7 \tfrac{\cot(\theta)a(\theta)^\prime}{a(\theta)} - \tfrac{1}{2} \tfrac{\cot(\theta) b(\theta)^\prime}{b(\theta)} + 7 \tfrac{a(\theta)^{\prime\prime}}{a(\theta)} - 14 \tfrac{a(\theta)^{\prime 2}}{a(\theta)^2} + \tfrac{1}{2} \tfrac{b(\theta)^{\prime\prime}}{b(\theta)} - \tfrac{1}{4} \tfrac{b(\theta)^{\prime 2}}{b(\theta)^2} + 7 \tfrac{a(\theta)^\prime b(\theta)^\prime}{a(\theta)b(\theta)} - 8 \tfrac{f(\theta)^{\prime 2}}{f(\theta)^2} \notag \\ &{} + \tfrac{49}{2} \tfrac{\lambda(\theta)^2 a(\theta)^{\prime 2}}{a(\theta)^2} + 21 \tfrac{\lambda(\theta)^2 a(\theta)^\prime f(\theta)^\prime}{a(\theta)f(\theta)} - 7 \tfrac{\lambda(\theta) a(\theta)^\prime \lambda(\theta)^\prime}{a(\theta)} + \tfrac{9}{2} \tfrac{\lambda(\theta)^2 f(\theta)^{\prime 2}}{f(\theta)^2} - 3\tfrac{\lambda(\theta) f(\theta)^\prime \lambda(\theta)^\prime}{f(\theta)} + \tfrac{1}{2} \lambda(\theta)^{\prime 2} = 0,
\label{eq:i1}
\end{align}
the internal Einstein equations transverse to the $\theta$-direction \eqref{eom5}
\begin{align}
& 7 \tfrac{\cot(\theta) a(\theta)^\prime}{a(\theta)} + \tfrac{3}{2} \tfrac{\cot(\theta) b(\theta)^\prime}{b(\theta)} + \tfrac{1}{2} \tfrac{b(\theta)^{\prime\prime}}{b(\theta)} - \tfrac{3}{4} \tfrac{b(\theta)^{\prime 2}}{b(\theta)^2} - \tfrac{7}{2} \tfrac{a(\theta)^\prime b(\theta)^\prime}{a(\theta)b(\theta)} + 2 - \tfrac{1}{16} \tfrac{F_0^2}{b(\theta)f(\theta)^{10}} - \tfrac{343}{16} \tfrac{\lambda(\theta)^2 a(\theta)^{\prime 2}}{a(\theta)^2} \notag \\ &{} - \tfrac{147}{8} \tfrac{\lambda(\theta)^2 a(\theta)^\prime f(\theta)^\prime}{a(\theta)f(\theta)} + \tfrac{49}{8} \tfrac{\lambda(\theta)a(\theta)^\prime \lambda(\theta)^\prime}{a(\theta)} - \tfrac{63}{16} \tfrac{\lambda(\theta)^2 f(\theta)^{\prime 2}}{f(\theta)^2} + \tfrac{21}{8} \tfrac{\lambda(\theta) f(\theta)^\prime \lambda(\theta)^\prime}{f(\theta)} - \tfrac{7}{16} \lambda^{\prime 2} - \tfrac{3}{8} \tfrac{\lambda(\theta)^2 F_0^2}{b(\theta)f(\theta)^{10}} \notag \\ &{} = \tfrac{7}{16} T \tfrac{\sqrt{b(\theta)} \delta(\theta)}{f(\theta)^3 \sin^2(\theta)},
\label{eq:i2}
\end{align}
a combination of the dilaton equation \eqref{eom2}, the Bianchi identity \eqref{eom1} and the transverse internal Einstein equations \eqref{eom5} such that all $F_0^2$-dependent terms cancel out which yields
\begin{align}
& 2 - \tfrac{7}{5} \tfrac{a(\theta)^\prime f(\theta)^\prime}{a(\theta)f(\theta)} - \tfrac{1}{10} \tfrac{b(\theta)^\prime f(\theta)^\prime}{b(\theta)f(\theta)}- \tfrac{7}{2} \tfrac{a(\theta)^\prime b(\theta)^\prime}{a(\theta)b(\theta)} + \tfrac{14}{5} \tfrac{\lambda(\theta)^2 a(\theta)^{\prime\prime}}{a(\theta)} + \tfrac{6}{5} \tfrac{\lambda(\theta)^2 f(\theta)^{\prime\prime}}{f(\theta)} - \tfrac{1}{5} \tfrac{f(\theta)^{\prime 2}}{f(\theta)^2} + \tfrac{2}{5} \tfrac{\cot(\theta) f(\theta)^\prime}{f(\theta)} \notag \\ &{} + 7 \tfrac{\cot(\theta)a(\theta)^\prime}{a(\theta)} - \tfrac{112}{5} \tfrac{\lambda(\theta)^2 a(\theta)^{\prime 2}}{a(\theta)^2} - \tfrac{6}{5} \tfrac{\lambda(\theta)^2 f(\theta)^{\prime 2}}{f(\theta)^2} + \tfrac{42}{5} \tfrac{\lambda(\theta) a(\theta)^\prime \lambda(\theta)^\prime}{a(\theta)} + \tfrac{12}{5} \tfrac{\lambda(\theta)f(\theta)^\prime \lambda(\theta)^\prime}{f(\theta)} - \tfrac{2}{5} \lambda(\theta)^{\prime 2} \notag \\ &{} - \tfrac{42}{5} \tfrac{\lambda(\theta)^2 a(\theta)^\prime f(\theta)^\prime}{a(\theta)f(\theta)} - \tfrac{3}{4} \tfrac{b(\theta)^{\prime 2}}{b(\theta)^2} + \tfrac{1}{5} \tfrac{f(\theta)^{\prime\prime}}{f(\theta)} + \tfrac{3}{2} \tfrac{\cot(\theta) b(\theta)^\prime}{b(\theta)} + \tfrac{28}{5} \tfrac{\cot(\theta) \lambda(\theta)^2 a(\theta)^\prime}{a(\theta)} + \tfrac{12}{5} \tfrac{\cot(\theta) \lambda(\theta)^2 f(\theta)^\prime}{f(\theta)} \notag \\ &{} + \tfrac{1}{5} \tfrac{\lambda(\theta) b(\theta)^\prime \lambda(\theta)^\prime}{b(\theta)} - \tfrac{4}{5} \cot(\theta) \lambda(\theta) \lambda(\theta)^\prime - \tfrac{7}{5} \tfrac{\lambda(\theta)^2 a(\theta)^\prime b(\theta)^\prime}{a(\theta)b(\theta)} - \tfrac{3}{5} \tfrac{\lambda(\theta)^2 b(\theta)^\prime f(\theta)^\prime}{b(\theta)f(\theta)} - \tfrac{2}{5} \lambda(\theta) \lambda(\theta)^{\prime\prime} \notag \\ &{} + \tfrac{1}{2} \tfrac{b(\theta)^{\prime\prime}}{b(\theta)} = \tfrac{2}{5} T \tfrac{\sqrt{b(\theta)} \delta(\theta)}{f(\theta)^3 \sin^2(\theta)} - \tfrac{2}{5} Q \tfrac{\lambda(\theta) \sqrt{b(\theta)} \delta(\theta)}{f(\theta)^3 \sin^2(\theta)},
\label{eq:i3}
\end{align}
a combination of the dilaton equation \eqref{eom2} and the internal Einstein equations \eqref{eom4} and \eqref{eom5} such that all $\lambda$-dependent terms cancel out which yields
\begin{align}
& 2 + 21\tfrac{a(\theta)^\prime f(\theta)^\prime}{a(\theta)f(\theta)} + \tfrac{3}{2} \tfrac{b(\theta)^\prime f(\theta)^\prime}{b(\theta)f(\theta)} + \tfrac{21}{2} \tfrac{a(\theta)^\prime b(\theta)^\prime}{a(\theta)b(\theta)} - 28 \tfrac{a(\theta)^{\prime 2}}{a(\theta)^2} - 13 \tfrac{f(\theta)^{\prime 2}}{f(\theta)^2} - 6 \tfrac{\cot(\theta) f(\theta)^\prime}{f(\theta)} - 7 \tfrac{\cot(\theta) a(\theta)^\prime}{a(\theta)} \notag \\ &{} - \tfrac{F_0^2}{b(\theta)f(\theta)^{10}} - 3 \tfrac{f(\theta)^{\prime\prime}}{f(\theta)} + 14 \tfrac{a(\theta)^{\prime\prime}}{a(\theta)} - \tfrac{5}{4} \tfrac{b(\theta)^{\prime 2}}{b(\theta)^2} + \tfrac{3}{2} \tfrac{b(\theta)^{\prime\prime}}{b(\theta)} + \tfrac{1}{2} \tfrac{\cot(\theta) b(\theta)^\prime}{b(\theta)} = T \tfrac{\sqrt{b(\theta)} \delta(\theta)}{f(\theta)^3 \sin^2(\theta)}
\label{eq:i4}
\end{align}
and finally a combination of the dilaton equation \eqref{eom2} as well as the external and internal Einstein equations \eqref{eom3} to \eqref{eom5} such that all $\lambda$-dependent terms cancel out which yields
\begin{align}
& 2 - 7\tfrac{a(\theta)^\prime f(\theta)^\prime}{a(\theta)f(\theta)} - \tfrac{1}{2} \tfrac{b(\theta)^\prime f(\theta)^\prime}{b(\theta)f(\theta)} - \tfrac{11}{2} \tfrac{a(\theta)^\prime b(\theta)^\prime}{a(\theta)b(\theta)} - 32 \tfrac{a(\theta)^{\prime 2}}{a(\theta)^2} - \tfrac{f(\theta)^{\prime 2}}{f(\theta)^2} + 2 \tfrac{\cot(\theta)f(\theta)^\prime}{f(\theta)} + 15 \tfrac{\cot(\theta)a(\theta)^\prime}{a(\theta)} + \tfrac{f(\theta)^{\prime\prime}}{f(\theta)} \notag \\ &{} + 4\tfrac{a(\theta)^{\prime\prime}}{a(\theta)} - 24 \tfrac{a(\theta)^2}{b(\theta)} - \tfrac{3}{4} \tfrac{b(\theta)^{\prime 2}}{b(\theta)^2} + \tfrac{1}{2} \tfrac{b(\theta)^{\prime\prime}}{b(\theta)} + \tfrac{3}{2} \tfrac{\cot(\theta) b(\theta)^\prime}{b(\theta)} = 0. \label{eq:i5}
\end{align}

\subsection{Leading order behaviour}

Let us now plug our ansatz \eqref{eq:aaa} into equations \eqref{eq:e1} to \eqref{eq:i5}. Note that since the different scalings $A,B,F,L$ are a priori unknown we can only trust the leading order (LO) terms of any equations (the sub-leading parts of the eoms may secretly contain more terms than we have written down, regardless of how many orders we consider in the expansion \eqref{eq:aaa}). For this section, we only need the leading order in the expansion \eqref{eq:aaa}, so we can disregard all terms $\sim \theta^\zeta$, $\sim \theta^\xi$, etc. for the moment. It is convenient to consider three different cases.

\subsubsection*{\bf Case 1: $ L > 0 $}

Plugging in our ansatz into \eqref{eq:e2}, we find that its possible LO terms are
\begin{equation}
-2 F (B-2+14A) \theta^{-2} + \frac{5}{4} \frac{F_0^2}{b_0 f_0^{10}} \theta^{-B-10F} = 0. \label{eq:b}
\end{equation}
Let us now check all possible values for $ F $. If $ 10 F > 2 - B $, we immediately get a contradiction, since then only the last term in above equation is of LO but cannot be zero. If $ 10 F = 2 - B $, all terms in above equation are of LO and we find
\begin{equation}
F_0^2 = -\frac{4}{25}(B-2)(B-2+14A) b_0 f_0^{10}. \label{eq:a}
\end{equation}
The LO of \eqref{eq:i3} gives
\begin{equation}
\frac{1}{25} (6 B^2+84AB-168A-24B-1) \theta^{-2} = 0
\end{equation}
and it follows that $A = -\tfrac{1}{84}\tfrac{6B^2-24B-1}{B-2} $. Note that $B \neq 2$, since otherwise \eqref{eq:i3} yields $1=0$. Plugging this into \eqref{eq:a}, we find
\begin{equation}
F_0^2 = -\frac{2}{3} b_0 f_0^{10}
\end{equation}
which is negative. Hence the possibility $ 10 F = 2 - B $ is excluded as well.

The last possibility we need to check is $ 10 F < 2 - B $. From the LO terms of \eqref{eq:b} we then get either $A = \tfrac{1}{7} - \tfrac{1}{14}B$ or $ F=0$. The first option plugged into \eqref{eq:i3} yields at LO
\begin{equation}
\theta^{-2} = 0
\end{equation}
and is therefore excluded. The second option plugged into \eqref{eq:i3} yields at LO
\begin{equation}
-\frac{1}{4} (-28A+14AB-4B+B^2) \theta^{-2} = 0
\end{equation}
which gives $ A = -\tfrac{1}{14} \tfrac{B(B-4)}{B-2} $. Note that by assumption $B < 2-10F$ and $F=0$ and hence $A$ is not singular. Plugging this into \eqref{eq:i1} and checking the LO terms gives
\begin{equation}
-\frac{1}{7} \frac{B(B-4)(2B^2-8B+7)}{(B-2)^2} \theta^{-2} = 0
\end{equation}
and yields four solutions $0,2-\tfrac{1}{2} \sqrt{2}, 2 + \tfrac{1}{2} \sqrt{2}, 4 $ for $ B$, where the latter two can be discarded because they violate our assumption $ 0 = 10 F < 2 - B $. With \eqref{eq:e3} also the other two solutions can be discarded, since its LO contribution evaluated for either $B=0$ or $B=2-\tfrac{1}{2} \sqrt{2}$ gives
\begin{equation}
L (2+2L) \theta^{L-2} = 0, \qquad (4L+\sqrt{2})(4L+7\sqrt{2})\theta^{L-2} = 0,
\end{equation}
respectively, which cannot be fulfilled for $L>0$.

\subsubsection*{Case 2: $ L = 0 $}

Plugging in our ansatz into \eqref{eq:e2}, we find
\begin{equation}
(4F -28AF -2BF +\frac{27}{4}\lambda_0^2 F^2+\frac{147}{4}\lambda_0^2A^2+\frac{63}{2}\lambda_0^2AF) \theta^{-2} + (\frac{5}{4} - \frac{1}{2} \lambda_0^2) \frac{F_0^2}{b_0 f_0^{10}} \theta^{-B-10F} = 0.
\end{equation}
and thus need to consider three possibilities: $ B=2-10F$, $B<2-10F$ and $B>2-10F$. The last possibility can easily be excluded by looking e.g. at \eqref{eq:e3} which then yields
\begin{equation}
\frac{F_0^2}{b_0 f_0^{10}} \theta^{-B-10F} = 0
\end{equation}
at LO, where the left-hand side cannot vanish. The first of the remaining two possibilities is excluded as follows:
We subtract the LO of \eqref{eq:e3} and \eqref{eq:i4} such that a linear equation for $B$ remains. This equation yields the condition $A \neq 0$ and can be solved for $B$ to obtain
\begin{equation}
B=2 + 2A + \frac{1}{7A} + \frac{3}{28} \frac{F_0^2}{b_0f_0^{10}A}.
\end{equation}
We then plug this result into the LO of \eqref{eq:i1} and \eqref{eq:i3} and combine the equations such that a quadratic equation for $A$ remains. The equation yields the condition $\lambda_0^2 \neq \frac{3}{2}$ and can be solved for $A$ to give
\begin{equation}
A = \pm \frac{\sqrt{14}}{112} \frac{\sqrt{b_0 (-3+2\lambda_0^2)(9\lambda_0^2 F_0^2 -6 F_0^2-8b_0f_0^{10}+12\lambda_0^2b_0f_0^{10})}}{b_0f_0^5(-3+2\lambda_0^2)}.
\end{equation}
We can then plug this into \eqref{eq:e3} and \eqref{eq:i3} and solve both equations for $F_0^2$ to find
\begin{equation}
F_0^2 = \frac{2b_0f_0^{10}(7\lambda_0^2-10)}{36-56\lambda_0^2+21\lambda_0^4}, \qquad F_0^2 = -\frac{2}{3}\frac{b_0f_0^{10}(54-73\lambda_0^2+28\lambda_0^4)}{48-82\lambda_0^2+35\lambda_0^4},
\end{equation}
where one can check that the denominators are not zero, since for those values of $\lambda_0$ \eqref{eq:i3} or \eqref{eq:e3} would not be satisfied. One can now combine the two equations such that $F_0^2$ cancels out and solve for $\lambda_0$. Plugging the solutions into one of the above equations, one then finds that $F_0^2$ is negative for all possible $\lambda_0$. Hence, our initial assumption $ B=2-10F$ has eventually led to a contradiction.

Now we check the second possibility $B<2-10F$. We first solve the LO of \eqref{eq:e2} for $B$ and find
\begin{equation}
B = \frac{1}{8F} (16F-112AF+27\lambda_0^2F^2+147\lambda_0^2A^2+126\lambda_0^2AF),
\end{equation}
unless $F=0$. (Note that if $F=0$ one gets $A=0$ from \eqref{eq:e2} and $B=0$ or $B=4$ from \eqref{eq:i1}, where the latter violates our assumption  $B<2-10F$. Hence, the only way to have $F=0$ is to have $A=B=F=L=0$ or in other words no divergent fields at all. This is a boundary condition with no source sitting at the pole, let us call it \emph{boundary condition 1}. We keep that in mind and proceed under the assumption that $F \neq 0$.) We then plug our finding for $B$ into the LO of \eqref{eq:e3} and solve for $A$ to find the solutions
\begin{equation}
A = -\frac{3}{7} F, \qquad A = -\frac{1}{21} \frac{F(-16+9\lambda_0^2)}{\lambda_0^2},
\end{equation}
where in the special case $\lambda_0^2 = 0$ one also finds $A = -\frac{3}{7} F$. Plugging in the first solution for $A$ into the LO of \eqref{eq:i1} gives a contradiction. Hence we continue with the second solution. Substituted into \eqref{eq:i2}, we can then solve for $F$ which yields
\begin{equation}
F = \pm \frac{3}{16} \lambda_0.
\end{equation}
From the LO of \eqref{eq:i1} we are finally left with an equation for $\lambda_0$. Solving this equation and testing the solutions for $\lambda_0$ in the LO of \eqref{eq:e1}, we find that only two solutions do not lead to a contradiction and that the sign of $F$ is fixed for consistency reasons:
\begin{equation}
\lambda_0 = \pm 1, \qquad F = - \frac{3}{16}.
\end{equation}
Thus we found \emph{boundary condition 2} with the properties
\begin{equation}
L=0, \qquad A = -\frac{1}{16}, \qquad B = \frac{7}{8}, \qquad F = -\frac{3}{16}, \qquad \lambda_0 = \pm 1
\end{equation}
and
\begin{equation}
\alpha = \frac{\lambda_0}{a_0^7f_0^3} \theta + \ldots, \label{alphabehavior2}
\end{equation}
where $\alpha(\theta)= \lambda(\theta) a(\theta)^{-7} f(\theta)^{-3}$.

\subsubsection*{Case 3: $ L < 0 $}

Considering the LO of \eqref{eq:i1} yields
\begin{equation}
A = \frac{1}{7}(L - 3 F).
\end{equation}
With this choice, also the LO of \eqref{eq:e2}, \eqref{eq:e3} and \eqref{eq:i2} are satisfied under the condition that $ B < 2 - 10F $.
The possible LO contributions of \eqref{eq:i4} and \eqref{eq:i5} then become
\begin{equation}
-\frac{1}{28}(-7B^2+772F^2-168F+8L^2-132FL+84L+84BF-42BL+28B) \theta^{-2} = 0 \label{eq:x1}
\end{equation}
and
\begin{equation}
\frac{1}{28}(28B- 104F-16L^2+68FL-60F^2+44L+52BF-22BL-7B^2) \theta^{-2} - 24\frac{a_0^2}{b_0} \theta^{\frac{2}{7}L-\frac{6}{7}F-B}=0. \label{eq:x2}
\end{equation}

There are two possibilities to solve \eqref{eq:x2} to LO: either $\frac{2}{7}L-\frac{6}{7}F-B = -2 $ or $\frac{2}{7}L-\frac{6}{7}F-B > -2 $. The first possibility is excluded, since it follows that $B = \frac{2}{7}L-\frac{6}{7}F + 2$ which can be plugged into \eqref{eq:x1} and \eqref{eq:x2} to find
\begin{equation}
L =  \frac{1}{4} \frac{32b_0F^2 + b_0 + 6a_0^2}{b_0 F},
\end{equation}
where one can check that $F \neq 0$. Plugging in our findings for $B$ and $L$ into \eqref{eq:x1}, one then verifies that its left-hand side is necessarily negative. Hence, \eqref{eq:x1} cannot be solved with this choice.

Now considering the second possibility $\frac{2}{7}L-\frac{6}{7}F-B > -2 $, we first add \eqref{eq:x1} and \eqref{eq:x2} such that the coefficient of $B^2$ is zero and solve for $B$ to find
\begin{equation}
B = \frac{2(8F-3L^2+25FL-104F^2-5L)}{8F-5L},
\end{equation}
where one can check that $F \neq \frac{5}{8}L$ and hence the denominator is not zero. Using our expressions for $A$ and $B$, we then find for the possible LO contributions of \eqref{eq:e1}
\begin{equation}
-2\frac{\lambda_0^2 F_0^2}{b_0f_0^{10}} \theta^{\frac{4(4F+L)(-8F+L)}{-8F+5L}-2} + \frac{32}{7} \frac{(-16F+3L)(16F^2-6FL+L^2)}{-8F+5L} \theta^{-2} = 0. \label{eq:x3}
\end{equation}

This now again yields two possibilities: either only the second term is of LO or both terms are of LO.

The first possibility yields $(-16F+3L)(16F^2-6FL+L^2)=0$ from \eqref{eq:x3} which gives one real and two complex solutions for $L$. Choosing the real solution $L= \frac {16}{3}F$, we can evaluate \eqref{eq:x1} to find $F= \pm \frac{3}{16}$. The sign of $F$ is then fixed to be negative by demanding that the previously found conditions $ B < 2 - 10F $ and $\frac{2}{7}L-\frac{6}{7}F-B > -2 $ hold. The result is then \emph{boundary condition 3}:
\begin{equation}
L = -1, \qquad A=-\frac{1}{16}, \qquad B=\frac{7}{8}, \qquad F=-\frac{3}{16}. \label{eq:bc3}
\end{equation}

The second possibility has the condition $(4F+L)(-8F+L) = 0$ with the solutions $F=-\frac{1}{4}L$ and $F=\frac{1}{8}L$. Plugging this into \eqref{eq:x1} and \eqref{eq:x3}, we find the two solutions
\begin{equation}
L = -\frac{\sqrt{42}}{6}, \qquad A=-\frac{5\sqrt{42}}{336}, \qquad B=2-\frac{\sqrt{42}}{8}, \qquad F=-\frac{\sqrt{42}}{48}, \qquad \lambda_0 = \pm  \frac{\sqrt{b_0}f_0^5}{\sqrt{3}F_0}, \label{eq:bc1}
\end{equation}
and
\begin{equation}
L = -\frac{\sqrt{3}}{6}, \qquad A=-\frac{\sqrt{3}}{24}, \qquad B=2-\frac{3\sqrt{3}}{4}, \qquad F=\frac{\sqrt{3}}{24}, \qquad \lambda_0 = \pm \frac{\sqrt{2 b_0}f_0^5}{\sqrt{3}F_0}, \label{eq:bc2}
\end{equation}
which we call \emph{boundary condition 4} and \emph{boundary condition 5}. One can check that both solutions satisfy the previously found conditions $ B < 2 - 10F $ and $\frac{2}{7}L-\frac{6}{7}F-B > -2 $ and are thus valid boundary conditions.

In total, we have found three boundary conditions \eqref{eq:bc3}, \eqref{eq:bc1} and \eqref{eq:bc2} with divergent $\lambda$ and two boundary conditions with regular $\lambda$. We will continue by determining the sources that they are compatible with.

\subsection{Sub-leading order behaviour}

For two of the boundary conditions found above, the sources they are compatible with depend on their sub-leading order behaviour, so we address this now.

\subsubsection*{Boundary condition 3: $ L = -1, A=-\frac{1}{16}, B=\frac{7}{8}, F=-\frac{3}{16} $}

We now consider our ansatz \eqref{eq:aaa} including next to leading order (NLO) and NNLO terms, e.g.
\begin{equation}
\e^{-A(\theta)} = a_0 \theta^A + a_1 \theta^{A+\zeta} + a_2 \theta^{A+\xi}
\end{equation}
and accordingly for the other functions $\e^{-2B}, \e^{-\frac{1}{4}\phi}, \lambda$. We then plug this into the eoms and use $ L = -1, A=-\frac{1}{16}, B=\frac{7}{8}, F=-\frac{3}{16} $ to find that the possible NLO contributions of \eqref{eq:i5} are
\begin{equation}
(2a_0b_1f_0 + a_0b_1f_0 \zeta + 22a_1b_0f_0 + 8a_1b_0f_0 \zeta + 2a_0b_0f_1 + 2a_0b_0f_1 \zeta) \zeta \theta^{-3+\zeta} + 48a_0^3f_0 \theta^{-2} = 0
\end{equation}
which cannot be solved to NLO unless $\zeta \le 1$. The NLO contributions of \eqref{eq:e1}, \eqref{eq:e3} and \eqref{eq:i4} are
\begin{align}
& 2 b_0 f_0^9 (3a_1f_0-a_0f_1) \zeta (\zeta+1) \theta^{-3+\zeta} + a_0 \lambda_0^2 F_0^2 \theta^{-2} - 36 a_0^3 f_0^{10} \theta^{-2} = 0, \\
& (-a_0 f_0 \lambda_1 + 7a_1 f_0 \lambda_0 + 3a_0 f_1 \lambda_0) \zeta (\zeta-1) \theta^{-4+\zeta} = 0, \\
& (-6a_0b_0f_1 + 3a_0b_1f_0 + 28a_1b_0f_0) \zeta (\zeta-1) \theta^{-3+\zeta} = 0.
\end{align}
Solving the equations for $\zeta < 1$ yields $a_1 = b_1 = f_1 = \lambda_1 = 0$ and hence we find that non-trivial NLO terms have $\zeta = 1$. Considering \eqref{eq:e1}, \eqref{eq:e3}, \eqref{eq:i4}, \eqref{eq:i5} at NNLO and assuming $\xi < 2$ yields analogous constraints and leads to $a_2 = b_2 = f_2 = \lambda_2 = 0$. Hence we can also set $\xi=2$.

Using $ \zeta=1 $, the NLO contributions of \eqref{eq:e1} and \eqref{eq:i5} become
\begin{align}
& 36 a_0^3 f_0^{10} + 4 a_0 b_0 f_0^9 f_1 - F_0^2 a_0 \lambda_0^2 - 12 a_1 b_0 f_0^{10} = 0, \\
&{} - 30 a_1 b_0 f_0 - 3 a_0 b_1 f_0 + 48 a_0^3 f_0 - 4 a_0 b_0 f_1 = 0.
\end{align}
All other eoms are identically satisfied at NLO and do therefore not give additional constraints. The NNLO contributions of \eqref{eq:e1}, \eqref{eq:e3}, \eqref{eq:i4} and \eqref{eq:i5} are
\begin{align}
& 120 a_1^2 b_0^2 f_0^{11} - 72 a_0 a_2 b_0^2 f_0^{11} + 24 a_0^2 b_0^2 f_0^{10} f_2 - 100 a_0 a_1 b_0^2 f_0^{10} f_1 + 12 a_0^2 b_0^2 f_0^9 f_1^2 + 18 a_0 a_1 b_0 b_1 f_0^{11} \notag \\ &{} + 144 a_0^3 a_1 b_0 f_0^{11} + 216 a_0^4 b_0 f_0^{10} f_1 - 6 a_0^2 b_0 b_1 f_0^{10} f_1 - 4 F_0^2 a_0^2 b_0 f_0 \lambda_0 \lambda_1 + 14 F_0^2 a_0^2 b_0 f_1 \lambda_0^2 \notag \\ &{} - 108 a_0^4 b_1 f_0^{11} + 3 F_0^2 a_0^2 b_1 f_0 \lambda_0^2 = 0, \\
& 4 a_0^2 b_0 f_0^2 \lambda_2 - 14 a_0 a_1 b_0 f_0^2 \lambda_1 + 14 a_1^2 b_0 f_0^2 \lambda_0 - 28 a_0 a_2 b_0 f_0^2 \lambda_0 - 12 a_0^2 b_0 f_0 f_2 \lambda_0 - 42 a_0 a_1 b_0 f_0 f_1 \lambda_0 \notag \\ &{} - 12 a_0^2 b_0 f_1^2 \lambda_0 - a_0^2 b_1 f_0^2 \lambda_1 + 7 a_0 a_1 b_1 f_0^2 \lambda_0 + 3 a_0^2 b_1 f_0 f_1 \lambda_0 = 0, \\
& 336 a_1^2 b_0^2 f_0^2 - 336 a_0 a_2 b_0^2 f_0^2 - 16 a_0^2 b_0^2 f_0^2 + 72 a_0^2 b_0^2 f_0 f_2 - 252 a_0 a_1 b_0^2 f_0 f_1 + 156 a_0^2 b_0^2 f_1^2 \notag \\ &{} - 126 a_0 a_1 b_0 b_1 f_0^2 - 36 a_0^2 b_0 b_2 f_0^2 - 18 a_0^2 b_0 b_1 f_0 f_1 + 15 a_0^2 b_1^2 f_0^2 = 0, \\
&{} - 6 a_0^2 b_0^2 f_0 f_2 - 38 a_0 a_2 b_0^2 f_0^2 - 38 a_0 a_1 b_0^2 f_0 f_1 + 47 a_1^2 b_0^2 f_0^2 - 3 a_0^2 b_0^2 f_1^2 - 2 a_0^2 b_0^2 f_0^2 + 72 a_0^4 b_0 f_0 f_1 \notag \\ &{} - 3 a_0^2 b_0 b_1 f_0 f_1 - 4 a_0^2 b_0 b_2 f_0^2  + 48 a_0^3 a_1 b_0 f_0^2 + 13 a_0 a_1 b_0 b_1 f_0^2 - 36 a_0^4 b_1 f_0^2 + 3 a_0^2 b_1^2 f_0^2 = 0.
\end{align}
With these equations solved, the NNLO of \eqref{eq:e2}, \eqref{eq:i1}, \eqref{eq:i2}, \eqref{eq:i3} each yield
\begin{equation}
(a_0 f_0 \lambda_1 - 7a_1f_0\lambda_0 - 3a_0 f_1\lambda_0)^2 = a_0^2 f_0^2. \label{eq:bc3constraint}
\end{equation}
The above seven equations determine seven of the eight coefficients $a_1,b_1,f_1,\lambda_1, a_2,b_2,f_2,\lambda_2$ which is sufficient to later determine the source terms. We are thus left with five free parameters $a_0,b_0,f_0,\lambda_0,\lambda_1$, one of which will still be fixed later by determining the charge of the source. One can check that higher order contributions of the eoms can consistently be solved by fixing the coefficients $a_n,b_n,f_n,\lambda_n$ with $n>2$, but do not yield additional constraints between the free parameters $a_0,b_0,f_0,\lambda_0,\lambda_1$.

It is also useful to compute $\alpha$. We find
\begin{equation}
\alpha = \frac{\lambda(x)}{a(x)^7 f(x)^3} = \frac{\lambda_0}{a_0^7 f_0^3} + \frac{a_0 f_0 \lambda_1 - 7a_1f_0\lambda_0  - 3a_0 f_1\lambda_0}{f_0^4 a_0^8}  \theta + \ldots \label{eq:alphabc3}
\end{equation}

\subsubsection*{Boundary condition 4: $ L = -\frac{\sqrt{42}}{6}, A=-\frac{5\sqrt{42}}{336}, B=2-\frac{\sqrt{42}}{8}, F=-\frac{\sqrt{42}}{48} $}

We start by showing that for these boundary conditions the sub-leading order terms have a structure involving four different scalings
\begin{eqnarray}
& \zeta = \frac{1}{16} \sqrt{42} + \frac{11}{336} \sqrt{42} \approx 0.617, \qquad \xi=-\frac{1}{12}\sqrt{42}+\frac{1}{12}\sqrt{330} \approx 0.974, & \notag \\ & \eta=\frac{1}{6} \sqrt{42} \approx 1.080, \qquad \kappa = 2 \label{eq:abc} &
\end{eqnarray}
and integer multiples thereof such that e.g.
\begin{align}
\e^{-A(\theta)} = & a_0 \theta^A + a_1 \theta^{A+\zeta} + a_2 \theta^{A+\xi} + a_3 \theta^{A+\eta} + a_4 \theta^{A+2\zeta} + a_5 \theta^{A+\xi+\zeta} + a_6 \theta^{A+\eta+\zeta} + a_7\theta^{A+3\zeta} \notag \\ &{} + a_8\theta^{A+2\xi}+ a_9\theta^{A+\kappa}+ a_{10}\theta^{A+\eta+\xi}+ a_{11}\theta^{A+2\eta} + \ldots \label{eq:abcd}
\end{align}
and accordingly for the other functions $\e^{-2B}, \e^{-\frac{1}{4}\phi}, \lambda$. In order to determine the source terms in the next section, we will only need the coefficients $a_3,b_3,f_3,\lambda_3$ and $a_{11},b_{11},f_{11},\lambda_{11}$.

Let us now suppose there is an additional term $a_\chi \theta^{A+\chi}$ somewhere in between $a_0 \theta^A$ and $a_{11} \theta^{A+2\eta}$ such that $0 < \chi < 2\eta$ and $\theta^{A+\chi} $ does not coincide with one of the already existent orders. Using $ L = -\frac{\sqrt{42}}{6}, A=-\frac{5\sqrt{42}}{336}, B=2-\frac{\sqrt{42}}{8}, F=-\frac{\sqrt{42}}{48} $, the contributions of \eqref{eq:e1}, \eqref{eq:e3}, \eqref{eq:i4} and \eqref{eq:i5} at that order are then
\begin{align}
&{} 63 F_0^2 a_0 b_\chi f_0 \lambda_0^2 - 84 F_0^2 a_0 b_0 f_0 \lambda_0 \lambda_\chi + 294 F_0^2 a_0 b_0 f_\chi \lambda_0^2 - 56 \sqrt{42} \chi a_\chi b_0^2 f_0^{11} + 14 \sqrt{42} \chi a_0 b_0^2 f_0^{10} f_\chi \notag \\ &{} + 42 a_0 b_0^2 f_0^{10} f_\chi - 7 a_0 b_0 b_\chi f_0^{11} -\sqrt{42} \chi a_0 b_0 b_\chi f_0^{11} + 84 \chi^2 a_0 b_0^2 f_0^{10} f_\chi - 252 \chi^2 a_\chi b_0^2 f_0^{11} = 0, \\
& (- 7 a_\chi f_0 \lambda_0 - 3 a_0 f_\chi \lambda_0 + a_0 f_0 \lambda_\chi)(6\chi-\sqrt{42}) = 0, \\
&{} 16 \sqrt{42} a_\chi b_0 f_0 - 168 \chi a_\chi b_0 f_0 + 36 \chi a_0 b_0 f_\chi - 2 \sqrt{42} a_0 b_0 f_\chi + 3 \sqrt{42} a_0 b_\chi f_0 - 18 \chi a_0 b_\chi f_0  = 0, \\
& 336 \chi a_\chi b_0 f_0 + 140 \sqrt{42} a_\chi b_0 f_0 + 14 \sqrt{42} a_0 b_0 f_\chi + 84 \chi a_0 b_0 f_\chi + 42 \chi a_0 b_\chi f_0 + 13 \sqrt{42} a_0 b_\chi f_0 = 0.
\end{align}
For $\chi \neq -\frac{1}{12}\sqrt{42}+\frac{1}{12}\sqrt{330}$ and $\chi \neq \frac{1}{6} \sqrt{42}$ (which is true by assumption) this yields the solution $a_\chi = b_\chi = f_\chi = \lambda_\chi = 0$. Hence, there cannot be an additional term with non-trivial coefficients in between $a_0 \theta^A$ and $a_{11} \theta^{A+2\eta}$ in \eqref{eq:abcd} (and accordingly for $\e^{-2B}, \e^{-\frac{1}{4}\phi}, \lambda$).

Now we compute the coefficients $a_3,b_3,f_3,\lambda_3$ and $a_{11},b_{11},f_{11},\lambda_{11}$. The subleading contributions of \eqref{eq:e1}, \eqref{eq:i4} and \eqref{eq:i5} at order $\sim \theta^\eta$ yield
\begin{align}
& 42 F_0^2 a_0 b_0 f_3 \lambda_0^2 + 34 a_0 b_0^2 f_0^{10} f_3 - 2 a_0 b_0 b_3 f_0^{11} - 98 a_3 b_0^2 f_0^{11} + 9 F_0^2 a_0 b_3 f_0 \lambda_0^2 \notag \\ &{} - 12 F_0^2 a_0 b_0 f_0 \lambda_0 \lambda_3 = 0, \\
&{} - 3 a_3 f_0 + a_0 f_3 = 0, \\
& 7 a_0 b_0 f_3 + 49 a_3 b_0 f_0 + 5 a_0 b_3 f_0 = 0,
\end{align}
where the other equations are identically satisfied.

At order $\sim \theta^{2\eta}$, \eqref{eq:e1}, \eqref{eq:e3}, \eqref{eq:i4} and \eqref{eq:i5} yield
\begin{align}
& 280 a_0 a_3 b_0^2 b_3 f_0^{12} + 9 a_0^2 b_0 b_3^2 f_0^{12} - 12 a_0^2 b_0^2 b_{11} f_0^{12} - 108 a_0^2 b_0^2 b_3 f_0^{11} f_3 + 192 a_0^2 b_0^3 f_0^{10} f_3^2 \notag \\
&{} - 45 F_0^2 a_0^2 b_3^2 f_0^2 \lambda_0^2 + 336 F_0^2 a_0^2 b_0^2 f_0 f_3 \lambda_0 \lambda_3 + 1736 a_3^2 b_0^3 f_0^{12} - 48 F_0^2 a_0^2 b_0^2 f_0^2 \lambda_0 \lambda_{11} \notag \\
&{}- 1568 a_0 a_3 b_0^3 f_0^{11} f_3 - 252 F_0^2 a_0^2 b_0 b_3 f_0 f_3 \lambda_0^2 + 168 F_0^2 a_0^2 b_0^2 f_0 f_{11} \lambda_0^2 + 24 F_0^2 a_0^2 b_0^2 f_0^2  \notag \\
&{}+ 36 F_0^2 a_0^2 b_0 b_{11} f_0^2 \lambda_0^2 - 672 F_0^2 a_0^2 b_0^2 f_3^2 \lambda_0^2 - 24 F_0^2 a_0^2 b_0^2 f_0^2 \lambda_3^2 + 72 F_0^2 a_0^2 b_0 b_3 f_0^2 \lambda_0 \lambda_3  \notag \\
&{}- 1120 a_0 a_{11} b_0^3 f_0^{12} + 360 a_0^2 b_0^3 f_0^{11} f_{11} = 0, \\
&{} - 98 a_0 a_3 b_0 f_0^{10} \lambda_3 + 28 a_0^2 b_0 f_0^{10} \lambda_{11} - 196 a_0 a_{11} b_0 f_0^{10} \lambda_0  + 98 a_3^2 b_0 f_0^{10} \lambda_0 + 49 a_0 a_3 b_3 f_0^{10} \lambda_0 \notag \\ &{} - 7 a_0^2 b_3 f_0^{10} \lambda_3 - 294 a_0 a_3 b_0 f_0^9 f_3 \lambda_0 - 84 a_0^2 b_0 f_0^9 f_{11} \lambda_0 + 21 a_0^2 b_3 f_0^9 f_3 \lambda_0 - 84 a_0^2 b_0 f_0^8 f_3^2 \lambda_0 \notag \\ &{} - 12 F_0^2 a_0^2 \lambda_0 = 0, \\
&{} - 1120 a_0 a_{11} b_0^2 f_0^{10} + 952 a_3^2 b_0^2 f_0^{10} - 210 a_0 a_3 b_0 b_3 f_0^{10} - 84 a_0^2 b_0 b_{11} f_0^{10} + 35 a_0^2 b_3^2 f_0^{10} \notag \\
&{}+ 280 a_0^2 b_0^2 f_0^9 f_{11}  - 1092 a_0 a_3 b_0^2 f_0^9 f_3 - 70 a_0^2 b_0 b_3 f_0^9 f_3 + 476 a_0^2 b_0^2 f_0^8 f_3^2 + 24 F_0^2 a_0^2 b_0 = 0, \\
&{} - 78 a_0^2 b_0 b_3 f_0 f_3 - 168 a_0^2 b_0^2 f_0 f_{11} - 1008 a_0 a_{11} b_0^2 f_0^2 + 81 a_0^2 b_3^2 f_0^2 - 84 a_0^2 b_0^2 f_3^2 + 1288 a_3^2 b_0^2 f_0^2 \notag \\ &{} - 980 a_0 a_3 b_0^2 f_0 f_3 + 350 a_0 a_3 b_0 b_3 f_0^2 - 108 a_0^2 b_0 b_{11} f_0^2 = 0.
\end{align}
\eqref{eq:i1} gives the equation
\begin{equation}
7 (7 a_3 f_0 \lambda_0 + 3a_0 f_3 \lambda_0 - a_0 f_0 \lambda_3)^2 =  6 a_0^2 b_0^2 f_0^2.
\end{equation}
If these equations are solved, all other eoms are then automatically satisfied at this order. Note that in none of the above equations there can be any interference with terms coming from other orders because of the numerical values \eqref{eq:abc} (e.g. $2\zeta \neq \eta$, $\zeta+\xi \neq \eta$, etc.).

Solving the equations yields
\begin{eqnarray}
& a_3 = \frac{3}{8} \frac{F_0^2 a_0 \lambda_0 \lambda_3}{b_0 f_0^{10}}, \qquad b_3 = - \frac{21}{4} \frac{F_0^2 \lambda_0 \lambda_3}{f_0^{10}}, \qquad f_3 = \frac{9}{8} \frac{F_0^2 \lambda_0 \lambda_3}{b_0 f_0^9}, \qquad \lambda_3 = \pm \frac{\sqrt{42}}{7}, & \notag \\ & a_{11} = - \frac{21}{64} \frac{F_0^2 a_0}{b_0 f_0^{10}}, \qquad b_{11} = \frac{141}{16} \frac{F_0^2}{f_0^{10}}, \qquad f_{11} = - \frac{387}{448} \frac{F_0^2}{b_0 f_0^9}, \qquad \lambda_{11} = 0, &
\end{eqnarray}
where $ \lambda_0 = \pm \frac{\sqrt{b_0} f_0^5}{\sqrt{3} F_0} $ as we have found above.

\subsection{Determining the sources}\label{sec:sourcedet}

The contributions to the sources from various fields originate from second derivatives of the fields that are physically sourced by the $\delta$-functions. In our case, we expect anti-D6-brane charge, so the term $\d F_2$ (which is $\sim \alpha''$) should be contributing. To account for the anti-D6-brane tension we need as well contributions from terms with second order derivatives on dilaton, and combinations of $\e^{2A}$ and $\e^{2B}$. To be able to make the calculations easy we can rewrite the equations of motion \eqref{eom1} to \eqref{eom5} so that they have total derivative terms on the left-hand side and the $\delta$-function with constant coefficient on the right-hand side. In this case the equations of motion, up to irrelevant terms, look like
\begin{equation}
\begin{split}
\left(e^{-\frac{3}{2}\phi-7A+B}\sin^2 \theta \left( e^{\frac{3}{4}\phi+7A} \lambda\right)'\right)' &= \ldots + Q \delta(\theta),\\
\left(e^{-\frac{7}{4}\phi+B}\sin^2\theta (e^\phi)'\right)' &= \ldots + \frac{3}{4}T \delta(\theta),\\
\left(e^{-\frac{3}{4}\phi-16A+B}\sin^2\theta (e^{16A})'\right)'&= \ldots + T \delta(\theta),\\
\left(e^{-\frac{3}{4}\phi-7A-B}\sin^2\theta (e^{7A+2B})'\right)'&= \ldots - \frac{7}{16} T \delta(\theta),\\
\left(e^{-\frac{3}{4}\phi}\sin^2\theta (e^B)'\right)'&= \ldots - \frac{7}{16} T \delta(\theta).
\end{split}
\end{equation}

By integrating these expressions we get
\begin{equation}\label{eq:inteeoms}
\begin{split}
e^{-\frac{3}{2}\phi-7A+B}\sin^2 \theta \left( e^{\frac{3}{4}\phi+7A} \lambda\right)' &= \int \ldots \d \theta + Q ,\\
e^{-\frac{7}{4}\phi+B}\sin^2\theta (e^\phi)' &= \int \ldots \d \theta + \frac{3}{4}T,\\
e^{-\frac{3}{4}\phi-16A+B}\sin^2\theta (e^{16A})'&= \int \ldots \d \theta + T ,\\
e^{-\frac{3}{4}\phi-7A-B}\sin^2\theta (e^{7A+2B})'&= \int \ldots \d \theta - \frac{7}{16} T ,\\
e^{-\frac{3}{4}\phi}\sin^2\theta (e^B)'&= \int \ldots \d \theta - \frac{7}{16} T .
\end{split}
\end{equation}
We can now insert the expansions into the left-hand side of these expressions and pick out the constant terms, which are the contributions to the source term.

\subsubsection*{Boundary condition 1: $ L = 0, A=0, B=0, F=0 $}

This does not give rise to any sources
\begin{equation}
Q = T = 0.
\end{equation}

\subsubsection*{Boundary condition 2: $ L = 0, A=-\frac{1}{16}, B=\frac{7}{8}, F=-\frac{3}{16} $}

This gives extremal sources
\begin{equation}\label{chargelambda2}
Q = \frac{\lambda_0 f_0^3}{\sqrt{b_0}} = \pm \frac{f_0^3}{\sqrt{b_0}},\quad T = \frac{f_0^3}{\sqrt{b_0}}.
\end{equation}

\subsubsection*{Boundary condition 3: $L = -1, A=-\frac{1}{16}, B=\frac{7}{8}, F=-\frac{3}{16} $}

This gives extremal sources, at NLO. One has to solve the NLO coefficients in terms of LO (see previous section) to find
\begin{equation}
Q = \pm \frac{f_0^3}{\sqrt{b_0}},\quad T = \frac{f_0^3}{\sqrt{b_0}}.
\end{equation}

\subsubsection*{Boundary condition 4: $ L = -\frac{\sqrt{42}}{6}, A=-\frac{5\sqrt{42}}{336}, B=2-\frac{\sqrt{42}}{8}, F=-\frac{\sqrt{42}}{48} $}

This gives inconsistent sources
\begin{equation}
Q = \pm \frac{f_0^3}{\sqrt{b_0}},\quad T = \left\{ \begin{array}{c} \frac{1}{3}\sqrt{\frac{14}{3}} \frac{f_0^3}{\sqrt{b_0}} \\ 5 \sqrt{\frac{2}{21}}\frac{f_0^3}{\sqrt{b_0}}\\ \frac{96-11\sqrt{42}}{21}\frac{f_0^3}{\sqrt{b_0}}\\ \frac{16-\sqrt{42}}{7}\frac{f_0^3}{\sqrt{b_0}}\end{array} \,. \right.
\end{equation}
Note that $Q$ here is found at NNNLO order and one has to solve NNNLO coefficients in terms of LO coefficients (see previous section) to get the above expression.

\subsubsection*{Boundary condition 5: $ L = -\frac{\sqrt{3}}{6}, A=-\frac{\sqrt{3}}{24}, B=2-\frac{3\sqrt{3}}{4}, F=\frac{\sqrt{3}}{24} $}

For this boundary condition we have LO terms in \eqref{eq:inteeoms} with positive powers of $\theta$, hence there cannot be any sub-leading order that gives constant terms, so
\begin{equation}
Q = T = 0.
\end{equation}

\bibliography{groups}

\bibliographystyle{utphysmodb}

\end{document}